 \documentclass[preprint,review,12pt]{elsarticle}

\usepackage{graphicx}

 \usepackage{amssymb}
 \usepackage{amsthm}
 \usepackage{color}
 \usepackage{fancybox}
 \usepackage{bm}
\newcommand{\bd}[1]{\mbox{\boldmath $#1$}}
\usepackage{hyperref}

\biboptions{comma,super}

\journal{Computers \& Fluids }

\begin{document}

\begin{frontmatter}



\title{Triple decomposition of velocity gradient tensor in homogeneous isotropic turbulence\tnoteref{label1}}
\tnotetext[label1]{\textcolor{red}{Accepted for publication. \url{https://doi.org/10.1016/j.compfluid.2019.104389}}}
\author[NU_ad]{Ryosuke Nagata}
\author[NU_ad]{Tomoaki Watanabe\corref{cor1}\fnref{fn1}}
\ead{watanabe.tomoaki@c.nagoya-u.jp}
\author[NU_ad]{Koji Nagata}
\author[L_ad]{Carlos B. da Silva}

\cortext[cor1]{Corresponding author (+81-052-789-3279)}
\address[NU_ad]{Department of Aerospace Engineering, Nagoya University, Nagoya 464-8603, Japan}

\address[L_ad]{Instituto Superior T\'ecnico, Universidade de Lisboa, Lisboa 1049-001, Portugal}

\begin{abstract}
The triple decomposition of a velocity gradient tensor is studied with direct numerical simulations of homogeneous isotropic turbulence, where the velocity gradient tensor $\nabla {\bd{u}}$ is decomposed into three components representing an irrotational straining motion $(\nabla\bd{u})_{\rm EL}$, a rigid-body rotation $(\nabla\bd{u})_{\rm RR}$, and a shearing motion  $(\nabla\bd{u})_{\rm SH}$. Strength of these motions can be quantified with the decomposed components. A procedure of the triple decomposition is proposed for three-dimensional flows, where the decomposition is applied in a basic reference frame identified by examining a finite number of reference frames obtained by three sequential rotational transformations of a Cartesian coordinate. Even though more than one basic reference frame may be available for the triple decomposition, the results of the decomposition depend little on the choice of basic reference frame. In homogeneous isotropic turbulence, regions with strong rigid-body rotations or straining motions are highly intermittent in space, while most flow regions exhibit moderately strong shearing motions in the absence of straining motions and rigid-body rotations. In the classical double decomposition, the velocity gradient tensor is decomposed into a rate-of-rotation tensor $\Omega_{ij}$ and a rate-of-strain tensor $S_{ij}$. Regions with large $\omega^2=2\Omega_{ij}\Omega_{ij}$ can be associated with rigid-body rotations and shearing motions while those with large $s^2=2S_{ij}S_{ij}$ can be associated with irrotational straining motions and shearing motions. Therefore, vortices with rigid-body rotations and shear layers in turbulence cannot be detected solely by thresholding $\omega$ or $s$ while they can be identified simply with $(\nabla\bd{u})_{\rm RR}$ and $(\nabla\bd{u})_{\rm SH}$ in the triple decomposition, respectively. The thickness of the shear layer detected in the triple decomposition is about 10 times of Kolmogorov scale, while the velocity parallel to the layer changes rapidly across the layer, in which the velocity difference across the shear layer is of the order of the root-mean-squared velocity fluctuation. 
\end{abstract}
\end{frontmatter}


\section{Introduction}\label{Introduction}
Turbulent flows appear in various scientific and engineering problems. In atmospheric science, turbulent boundary layers formed in the atmosphere are important in many aspects~\cite{mahrt2014stably} because turbulence affects wind speed, temperature, heat transfer, and dispersion of pollutants. In engineering applications, the nature of turbulence is often utilized for improving the performance of equipment. For examples, chemical processes in reactors and propulsion systems often rely on the mixing of fluids enhanced by turbulence.~\cite{cumpsty2015jet,stankiewicz2000process} \par
A velocity gradient tensor $\nabla {\bd{u}}$ is a useful quantity to characterize a turbulent fluid motion at a point with respect to surrounding fluids. Herein, a component of a second-order tensor is denoted with subscripts, e.g., $(\nabla {\bd{u}})_{ij}=\partial u_{i}/\partial x_{j}$. In incompressible flows, the velocity gradient tensor can be decomposed into the rate-of-strain tensor $S_{ij} = (\partial u_i/\partial x_j + \partial u_j/\partial x_i )/2$ and the rate-of-rotation tensor $\Omega_{ij} = (\partial u_i/\partial x_j - \partial u_j/\partial x_i)/2$, separately.~\cite{pope2000turbulent} The double decomposition $(\nabla {\bd{u}})_{ij}=S_{ij}+\Omega_{ij}$ has been proven to be useful for investigating turbulence. A vorticity vector ${\bd{\omega}}=\nabla\times{\bd{u}}$ is expressed by the components of $\Omega_{ij}$. The rate-of-strain tensor can be used for expressing the viscous dissipation rate of kinetic energy as $\varepsilon=2\nu S_{ij}S_{ij}$ ($\nu$: kinematic viscosity), which is an essential quantity in turbulence theories and models. \par
\begin{figure}[t]
  \begin{center}
    \includegraphics[width=0.6\linewidth,keepaspectratio]{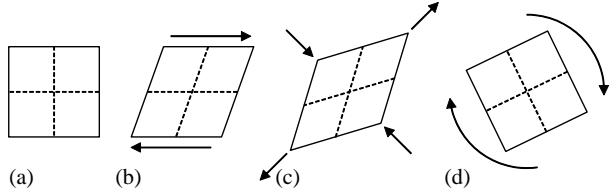}
    \caption{Sketches of fluid motions around a square fluid element extracted by the triple decomposition: (a) square fluid element; (b) shearing motion; (c) irrotational straining motion; (d) rigid-body rotation. }
    \label{Fig_TDM}
  \end{center}
\end{figure}

Turbulent flows are often studied in terms of turbulent structures, which can be identified with their characteristic patterns in flow visualization. A well-known structure that can be found in most turbulent flows is a small-scale tube-like vortex. It can be identified with quantities related to the velocity gradient tensor. For example, vortices can be found in a region with high enstrophy.~\cite{jimenez1998characteristics,da2011intense,jahanbakhshi2015baroclinic} Various methods have been proposed for identifying vortices in turbulence. However, its definition is not unique, and many identification schemes for vortices have been proposed in previous studies.~\cite{hunt1988eddies,chong1990general,jeong1995identification,haller2005objective,kolavr2007vortex} The methods based on the velocity gradient tensor mentioned above are local criteria applied at each flow point. It has been reported that vortex identification based on $\omega^{2}/2$ cannot distinguish a swirling motion associated with a vortex and a shearing motion.~\cite{kolavr2007vortex} For example, a parallel shear flow, such as the initial state of a mixing layer, exhibits a non-zero vorticity arising from the shear that is different from a vortex. The same problem is encountered in vortex identification in turbulent flows because a flow region with a strong shear can appear along with vortices. The triple decomposition of the velocity gradient tensor~\cite{kolavr2007vortex} was proposed for overcoming the shortcoming of the double decomposition. The triple decomposition decomposes the velocity gradient tensor into three tensors associated with an irrotational straining motion, a rigid-body rotation, and a shearing motion. Figure~\ref{Fig_TDM} illustrates a square fluid element under these motions in two dimensions.~\cite{kolavr2007vortex} A shearing motion is expressed by the parallel relative motion, where the planes parallel to the motion are not deformed and rotated. In the triple decomposition, the rigid-body rotation expresses a swirling motion that is not contributed by shearing and irrotational straining motions. Thus, the tensor of the rigid-body rotation can be used for identifying vortices. \par
Many studies on vortices in turbulent flows have been performed. Recently, turbulent flows have been studied with special focus on a flow region with strong shear. A thin internal shear layer with a cluster of strong vortex tubes was found in isotropic turbulence at high Reynolds number.~\cite{ishihara2013thin} An internal shear layer was shown to exist in turbulent boundary layers.~\cite{robinson1991coherent} The internal shear layer separates a large-scale flow structure from a flow region associated with different flow structures.~\cite{meinhart1995existence} Investigation of the internal shear layer requires a robust identification scheme similar to vortex identifications. Hence, Eisma et al.~\cite{eisma2015interfaces} utilized the triple decomposition, where the internal shear layer is detected with a tensor representing a shearing motion. The triple decomposition is expected to be applied to different flows to reveal the nature of internal shear layers that can exist in various turbulent flows. \par
The triple decomposition was originally proposed to decompose the velocity gradient tensor in three-dimensional flows, i.e., $(\nabla {\bd{u}})_{ij}$ with $i,j=1,2,$ and $3$. However, the procedure for applying the triple decomposition of the full velocity gradient tensor has not been proposed in previous studies. The triple decomposition requires identification of a basic reference frame where the decomposition of $(\nabla {\bd{u}})_{ij}$ can extract the tensors representing the three motions. This basic reference frame in three dimensional cases cannot be analytically obtained. Therefore, the triple decomposition has been applied to the velocity gradient tensor defined on a two-dimensional plane in three-dimensional flows, $(\nabla {\bd{u}})_{ij}$ with $i,j=1$ and $2$, because the basic reference frame is analytically obtained from two-dimensional components of $(\nabla {\bd{u}})_{ij}$.~\cite{kolavr2007vortex,maciel2012method,eisma2015interfaces} Turbulent flows comprise complex three-dimensional fluid motions; vortices and internal shear layers are extended in a three-dimensional space, in which their characteristics in the three-dimensional space are of high interest. In the present study, the triple decomposition is tested in three-dimensional flows using a direct numerical simulation (DNS) database of incompressible, forced, and homogeneous isotropic turbulence, which is a canonical flow studied by turbulence researchers. The general numerical procedure of the triple decomposition in a three-dimensional flow field is presented for applying the decomposition to the full velocity gradient tensor. Once the basic reference frame is determined by the proposed method, the triple decomposition is applied to the velocity gradient tensor with the formula presented by Kol{\'a}{\v{r}}.~\cite{kolavr2007vortex} The proposed procedure is tested with the DNS database in terms of computation time and flow fields described with the decomposed tensor. The results of the triple decomposition are discussed in comparison with the classical double decomposition. The triple decomposition is also used for detecting the region with strong shear. \par
\section{Triple decomposition of velocity gradient tensor}\label{Sec_Algorithm}
The triple decomposition~\cite{kolavr2007vortex} decomposes the velocity gradient tensor into the shearing motion tensor $\left(\nabla \bd{u} \right)_{\rm SH}$ and the residual tensor defined as $(\nabla \bd{u})_{\rm RES}=\nabla \bd{u}-\left(\nabla \bd{u} \right)_{\rm SH}$. Subsequently, $(\nabla \bd{u})_{\rm RES}$ is further decomposed into two components associated with the irrotational straining motion (elongation) $\left(\nabla \bd{u} \right)_{\rm EL}$ and rigid-body rotation $\left(\nabla \bd{u} \right)_{\rm RR}$. The subscripts EL, RR, and SH denote elongation, rigid-body rotation, and shear, respectively. Thus, the triple decomposition decomposes the velocity gradient tensor into three components as $\nabla \bd{u} = \left(\nabla \bd{u} \right)_{\rm EL} + \left(\nabla \bd{u} \right)_{\rm RR} + \left(\nabla \bd{u} \right)_{\rm SH}$. 
A contribution of the shearing motion is given by $\left(\nabla \bd{u} \right)_{\rm SH}$ that is a purely asymmetric tensor satisfying $\{\left(\nabla \bd{u} \right)_{\rm SH}\}_{ij}\{\left(\nabla \bd{u} \right)_{\rm SH}\}_{ji}=0$ (no summation). $\left(\nabla \bd{u} \right)_{\rm EL}$ and $\left(\nabla \bd{u} \right)_{\rm RR}$ are the symmetric and antisymmetric parts of $(\nabla \bd{u})_{\rm RES}$, respectively. An interaction scalar~\cite{kolavr2007vortex} $I$ is defined with $S_{ij}$ and $\Omega_{ij}$ by $I\equiv|S_{12} \Omega_{12}| + |S_{23} \Omega_{23}| + |S_{31} \Omega_{31}|$. 
The basic reference frame is defined such that $I$ assumes the maximum value among all reference frames.~\cite{kolavr2007vortex} \par
The following algorithm is used for applying the triple decomposition in three-dimensional flows in this study. 
The basic reference frame is identified from the reference frames obtained with the following rotational transformation tensor ${\bd{Q}}(\alpha,\beta, \gamma)$ applied to a Cartesian coordinate ${\bd{x}}=(x,y,z)$:
\begin{eqnarray}
{\bd Q} = \left(
\begin{array}{ccc}
\cos\alpha \cos\beta \cos\gamma - \sin\alpha \sin\gamma & \sin\alpha \cos\beta \cos\gamma + \cos\alpha \sin\gamma & -\sin\beta \cos\gamma \\
-\cos\alpha \cos\beta \sin\gamma - \sin\alpha \cos\gamma & -\sin\alpha \cos\beta \sin\gamma + \cos\alpha \cos\gamma & \sin\beta \sin\gamma \\
\cos\alpha \sin\beta & \sin\alpha \sin\beta & \cos\beta
\label{Eq_Q}
\end{array}
\right).
\end{eqnarray}
This is obtained as three sequential rotational transformations with angles $\alpha$, $\beta$, and $\gamma$ in the ranges of $0^{\circ}\leq \alpha \leq 180^{\circ}$, $0^{\circ}\leq \beta  \leq 180^{\circ}$, and $0^{\circ}\leq \gamma \leq 90^{\circ}$. The velocity gradient tensor expressed in the rotated reference frame ${\bd{x}}^{*}={\bd{Qx}}$ is obtained as $\left( \nabla\bd{u} \right)^{*} = {\bd Q}\left(\nabla \bd{u} \right){\bd Q^{T}}$, where the superscript $*$ represents a quantity evaluated in the rotated reference frame. A finite number of reference frames are specified with integers $(l,m,n)$ as $(\alpha, \beta, \gamma)=(l\Delta, m\Delta, n\Delta)$, where $0\leq l \leq 180^{\circ}/\Delta$, $0\leq m \leq 180^{\circ}/\Delta$, and $0\leq n \leq 90^{\circ}/\Delta$, and $\Delta$ is a small angle in units of degrees. For each reference frame with $(\alpha, \beta, \gamma)=(l\Delta, m\Delta, n\Delta )$, a rotational transformation is applied to the velocity gradient tensor, and $I^{*}$ is computed in the rotated reference frame. Subsequently, the basic reference frame can be determined with $(\alpha, \beta, \gamma)$ that yields the maximum value of $I^{*}$ among the reference frames considered. Apparently, a smaller $\Delta$ value results in a more accurate choice of the basic reference frame although it increases the computational cost. \par
Once the basic reference frame is determined, the triple decomposition is applied for the velocity gradient tensor $\left( \nabla\bd{u} \right)^{*}$ in the basic reference frame, yielding $(\nabla \bd{u})^{*} = (\nabla \bd{u})_{\rm RES}^{*} + (\nabla \bd{u})_{\rm SH}^{*}$. Here, $(\nabla \bd{u})_{\rm RES}^{*}$ is given by the following equation:~\cite{kolavr2007vortex} 
\begin{eqnarray}
\left\{ (\nabla \bd{u})_{\rm RES}^{*} \right\}_{ij} = {\mathrm{sgn}} \left[ (\nabla \bd{u})^{*}_{ij}\right] \min \left[ |(\nabla \bd{u})^{*}_{ij}|,|(\nabla \bd{u})^{*}_{ji}| \right],
\label{Eq_RES_BRF}
\end{eqnarray}
where ${\mathrm{sgn}}(x)$ is the sign function that yields ${\mathrm{sgn}}(x)=1$ for $x>0$, ${\mathrm{sgn}}(x)=0$ for $x=0$, and ${\mathrm{sgn}}(x)=-1$ for $x<0$. $(\nabla \bd{u})_{\rm SH}^{*}$ can be obtained by subtracting $(\nabla \bd{u})_{\rm RES}^{*}$ from $(\nabla \bd{u})^{*}$. In Eq.~(\ref{Eq_RES_BRF}), the diagonal part can be written as $\left\{ (\nabla \bd{u})_{\rm RES}^{*} \right\}_{ii}=(\nabla \bd{u})^{*}_{ii}$ (no summation for $i$). Equation~(\ref{Eq_RES_BRF}) yields $(\nabla \bd{u})_{\rm SH}$ in a purely asymmetric tensor form with $\left\{(\nabla \bd{u})_{\rm SH}^{*}\right\}_{ij}\left\{(\nabla \bd{u})_{\rm SH}^{*}\right\}_{ji} = 0$ (no summation). Finally, $(\nabla \bd{u})_{\rm RES}^{*}$ can be further decomposed into symmetric and antisymmetric parts that are associated with the elongation $\left(\nabla \bd{u} \right)_{\rm EL}^{*}$  and the rigid-body rotation $\left(\nabla \bd{u} \right)_{\rm RR}^{*}$, respectively: 
$\{(\nabla \bd{u} )_{\rm EL}^{*}\}_{ij}
=  [\{(\nabla \bd{u})_{\rm RES}^{*}\}_{ij}
   +\{(\nabla \bd{u})_{\rm RES}^{*}\}_{ji}]/2$ and 
$\{(\nabla \bd{u} )_{\rm RR}^{*}\}_{ij}
=  [\{(\nabla \bd{u})_{\rm RES}^{*}\}_{ij}
   -\{(\nabla \bd{u})_{\rm RES}^{*}\}_{ji}]/2$.
The results of the decomposition are valid for arbitrary reference frames obtained with the orthogonal transformation of the basic reference frame. Therefore, $\left(\nabla \bd{u} \right)_{\rm EL}$, $\left(\nabla \bd{u} \right)_{\rm RR}$, and $\left(\nabla \bd{u} \right)_{\rm SH}$ in the laboratory reference frame can be obtained by applying the inverse tensor of the rotational transformation tensor. It is noteworthy that the rotational transformation tensor is an orthogonal tensor that satisfies ${\bd Q^{T}}={\bd{Q}}^{-1}$. Therefore, $\left(\nabla \bd{u} \right)_{\rm EL}$ can be obtained by $\left( \nabla\bd{u} \right)_{\rm EL} = {\bd Q}^{T}\left(\nabla \bd{u} \right)_{\rm EL}^{*}{\bd Q}$. Similarly, $\left( \nabla\bd{u} \right)_{\rm RR}$ and $\left( \nabla\bd{u} \right)_{\rm SH}$ can also be obtained from $\left( \nabla\bd{u} \right)_{\rm RR}^{*}$ and $\left( \nabla\bd{u} \right)_{\rm SH}^{*}$. The triple decomposition is applied to the velocity gradient tensor defined at one point in a flow. The procedure above can be repeated for all points of the flow to obtain the spatial distributions of $\left( \nabla\bd{u} \right)_{\rm EL}$, $\left( \nabla\bd{u} \right)_{\rm RR}$, and $\left( \nabla\bd{u} \right)_{\rm SH}$. \par
 $\left( \nabla\bd{u} \right)_{\rm SH}$ can be decomposed into symmetric and antisymmetric parts as
$(S_{\mathrm{SH}})_{ij}=[\{( \nabla\bd{u} )_{\rm SH}\}_{ij}+\{( \nabla\bd{u} )_{\rm SH}\}_{ji}]/2$ and 
$(\Omega_{\mathrm{SH}})_{ij}=[\{( \nabla\bd{u} )_{\rm SH}\}_{ij}-\{( \nabla\bd{u} )_{\rm SH}\}_{ji}]/2$. 
The following symbols are introduced for representing the norm of the tensors:
$\omega^2={2\Omega_{ij}\Omega_{ij}}$; 
$\omega_{\mathrm{RR}}^2=2\{( \nabla\bd{u} )_{\rm RR}\}_{ij}\{( \nabla\bd{u} )_{\rm RR}\}_{ij}$; 
$\omega_{\mathrm{SH}}^2=2( \Omega_{\rm SH} )_{ij}( \Omega_{\rm SH} )_{ij}$; 
$s^2=2S_{ij}S_{ij}$; 
$s_{\mathrm{EL}}^2=2\{( \nabla\bd{u} )_{\rm EL}\}_{ij}\{( \nabla\bd{u} )_{\rm EL}\}_{ij}$; 
$s_{\mathrm{SH}}^2=2( S_{\rm SH} )_{ij}( S_{\rm SH} )_{ij}$. 
$\omega_{\mathrm{RR}}^2$ can be used for vortex identification.~\cite{kolavr2007vortex} Similarly, $s_{\mathrm{EL}}^2$ quantifies the strength of the irrotational straining motion. Because of the relation $\omega_{\mathrm{SH}}=s_{\mathrm{SH}}$, the strength of the shearing motion is expressed by either $\omega_{\mathrm{SH}}$ or $s_{\mathrm{SH}}$. The results of the triple decomposition are presented with $\omega_{\mathrm{RR}}$, $\omega_{\mathrm{SH}}$, $s_{\mathrm{EL}}$, and $s_{\mathrm{SH}}$. \par
\section{Direct Numerical Simulations of Homogeneous Isotropic Turbulence}\label{Sec_Direct}
The triple decomposition is applied to the DNS database of forced homogeneous isotropic turbulence. The DNS was performed with a classical pseudo-spectral method that solves the incompressible Navier--Stokes equations, and the details of the DNS have been described in previous studies.~\cite{da2008invariants,teixeira2012turbulence,valente2014effect,elsinga2017scaling} The statistically stationary turbulence is simulated with an artificial forcing that is delta correlated in time and uncorrelated with the velocity field.~\cite{alvelius1999random} Herein, the velocity components in the $x$, $y$, and $z$ directions are denoted by $u$, $v$, and $w$, respectively. The computational domain is a periodic box with a side length of $2\pi$ and is represented with $N^3$ grid points. \par
\begin{table}
  \begin{center}
  \caption{DNS of forced homogeneous isotropic turbulence. }
  \label{Table_DNS}
  \begingroup
  \begin{tabular}{cccccccccccc}
  \hline

  & Case
  & $N$
  & $\nu$
  & $u_{\mathrm{rms}}$
  & $\langle \varepsilon\rangle$
  & $Re_0$
  & $Re_\lambda$
  & $L_u/\eta$
  & $\lambda/\eta$
  & $\Delta_{xyz}/\eta$\\
  \hline

  & N128
  & 128
  & $0.025$
  & $2.96$
  & $10.1$
  & $101$
  & $49$
  & $27$
  & $14$
  & $1.4$\\


  & N512
  & 512
  & $0.00375$
  & $2.65$
  & $10.2$
  & $933$
  & $177$
  & $140$
  & $26 $
  & $1.4$\\

  \hline
  \end{tabular}
  \endgroup
  \end{center}
\end{table}
The two DNS databases summarized in Tab.~\ref{Table_DNS} are analyzed using a workstation (CPU: Intel Xeon E5-1620 v3, 3.50 GHz). In homogeneous isotropic turbulence, an average $\langle\,\,\,\rangle$ of physical variables is computed as the spatial average in the computational domain. Table~\ref{Table_DNS} shows the kinematic viscosity $\nu$, root-mean-squared velocity fluctuation $u_{\mathrm{rms}}$, and kinetic energy dissipation rate $\langle\varepsilon\rangle$ in arbitrary units, where the subscript rms denotes a root-mean-squared fluctuation $f_{\mathrm{rms}}=\sqrt{\langle f^2\rangle - \langle f\rangle^2}$. Table~\ref{Table_DNS} shows the ratio among the integral length scale $L_u$, Taylor microscale $\lambda$, and Kolmogorov length scale $\eta$, where $L_u$ is obtained by integrating the longitudinal autocorrelation function;  $\lambda$ and $\eta$ are defined as $\lambda=u_{\mathrm{rms}}/(\partial u/\partial x)_{\mathrm{rms}}$ and $\eta=(\nu/\langle\varepsilon\rangle^3)^{1/4}$, respectively. Table~\ref{Table_DNS} also shows the grid size $\Delta_{xyz}=2\pi/N$ divided by $\eta$. The Kolmogorov length scale is often considered as the characteristic length scale of the smallest turbulent motions, and the grid size is small enough to perform direct numerical simulations.~\cite{pope2000turbulent} Table~\ref{Table_DNS} also shows the Reynolds number based on the integral scale $Re_{0}=u_{\mathrm{rms}}L_u/\nu$ and turbulent Reynolds number $Re_{\lambda}=u_{\mathrm{rms}}\lambda/\nu$. \par
\begin{figure}[t]
  \begin{center}
    \includegraphics[width=0.9\linewidth,keepaspectratio]{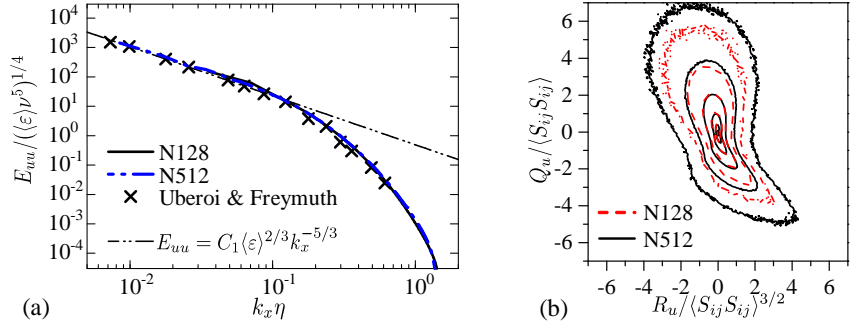}
    \caption{(a) Normalized one-dimensional energy spectra $E_{uu}/(\langle\varepsilon\rangle\nu^{5})^{1/4}$, plotted against $k_{x}\eta$, where $k_{x}$ is the wavenumber in the $x$ direction. The present results are compared with the experimental results of turbulent wake by Uberoi \& Freymuth~\cite{uberoi1970turbulent} and $E_{uu}=C_{1}\langle \varepsilon \rangle^{2/3}k_{x}^{-5/3}$ with $C_{1}=0.49$. (b) Isolines of the joint PDF of $Q_u$ and $R_u$ normalized by $\langle S_{ij}S_{ij}\rangle$. The contour levels are $10^{-4}$, $10^{-3}$, $10^{-2}$, $10^{-1}$, $10^{0}$, and $10^{1}$ (from outside to center of the figure). }
    \label{E11_rev}
  \end{center}
\end{figure}
Figure~\ref{E11_rev}(a) shows the one-dimensional energy spectra of $u$, $E_{uu}$, normalized by $(\langle\varepsilon\rangle\nu^{5})^{1/4}$. The present DNS results agree well with the experimental results in a turbulent wake~\cite{uberoi1970turbulent} and the relation $E_{uu}=C_{1}\langle \varepsilon \rangle^{2/3}k_{x}^{-5/3}$  for the inertial subrange. Figure~\ref{E11_rev}(b) shows the joint probability density function (PDF) of the second and third invariants of $\nabla \bd{u}$ that are defined as  $Q_{u}=-(\nabla \bd{u})_{ij}(\nabla \bd{u})_{ji}/2$ and $R_{u}=-(\nabla \bd{u})_{ij}(\nabla \bd{u})_{jk}(\nabla \bd{u})_{ki}/3$, respectively. The joint PDF exhibits the well-known teardrop shape in consistent with previous studies of turbulent flows.~\cite{davidson2004turbulence} A larger probability for $|Q_{u}|\gg 0$ or $|R_{u}|\gg 0$ can be found for the higher Reynolds number, where this tendency can be explained by larger flatness of the velocity derivative for a higher Reynolds number. 
\par
\section{Results and Discussion}\label{Sec_Results}
\subsection{Dependence of triple decomposition on choice of basic reference frame}
\begin{figure}[t]
  \begin{center}
    \includegraphics[width=1\linewidth,keepaspectratio]{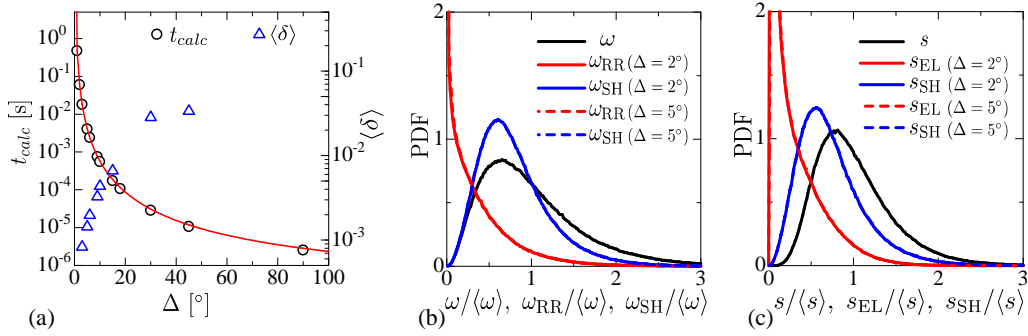}
    \caption{(a) Computation (CPU) time $t_{calc}$ of the triple decomposition applied at one grid point and $\langle \delta \rangle=\langle |\omega_{\rm SH}(\Delta)-\omega_{\rm SH}(2^{\circ})|/\omega_{\rm SH}(2^{\circ})\rangle$  as a function of $\Delta$. The red line represents $t_{calc}\propto (180^{\circ}/\Delta+1)^2 (90^{\circ}/\Delta+1)$. PDFs of (b) $\omega$, $\omega_{\rm RR}$, and $\omega_{\rm SH}$ and (c) $s$, $s_{\rm EL}$, and $s_{\rm SH}$. The results of the triple decomposition are shown for $\Delta = 2^{\circ}$ or $5^{\circ}$. These results are obtained from N128. }
    \label{Calc_time}
  \end{center}
\end{figure}

The basic reference frame obtained by the present procedure depends on the parameter $\Delta$ introduced in Sec.~\ref{Sec_Algorithm}. The triple decomposition is tested for $1^{\circ}\leq \Delta \leq90^{\circ}$ for N128. It is expected that a smaller value of $\Delta$ enables us to more accurately obtain the basic reference frame, where the interaction scalar $I$ assumes the largest value among all possible reference frames. However, the computational time of the triple decomposition increases as $\Delta$ becomes smaller. Figure~\ref{Calc_time}(a) shows the computation time $t_{calc}$ (CPU time) required for applying the triple decomposition at one grid point for various values of $\Delta$. The red line in the figure is $t_{calc}=C(180^{\circ}/\Delta+1)^2(90^{\circ}/\Delta+1)$, where $(180^{\circ}/\Delta+1)^2(90^{\circ}/\Delta+1)$ is the number of reference frames considered to obtain the basic reference frame, and the constant $C$ is obtained as $C=t_{calc}/[(180^{\circ}/\Delta+1)^2(90^{\circ}/\Delta+1)]$ with $t_{calc}$ measured for $\Delta=1^{\circ}$. The computation time increases rapidly as $\Delta$ decreases, where $t_{calc}\propto(180^{\circ}/\Delta+1)^2(90^{\circ}/\Delta+1)$ agrees well with the measured computation time. 
Because the computation time decreases rapidly with $\Delta$, increasing $\Delta$ from $1^{\circ}$ reduces the computation time significantly. \par
The triple decomposition is applied to the whole computational domain in N128 for $2^{\circ}\leq \Delta \leq45^{\circ}$. Here, the minimum value $\Delta=2^{\circ}$ is restricted by the computational time. The dependence of the decomposition on $\Delta$ is examined by $\omega_{\rm SH}$ obtained as a function of $\Delta$ at each computational grid point, where the relative deviation $\omega_{\rm SH}(\Delta)$ from $\omega_{\rm SH}(2^{\circ})$ is defined as $\delta=(|\omega_{\rm SH}(\Delta)-\omega_{\rm SH}(2^{\circ})|)/\omega_{\rm SH}(2^{\circ})$. The averaged value $\langle \delta \rangle$ is plotted against $\Delta$ in Fig.~\ref{Calc_time}(a), where $\langle \delta \rangle$ becomes small as $\Delta$ decreases. $\langle \delta \rangle$ is less than $1\%$ for $\Delta\leq 15^{\circ}$, and the results of the decomposition depend little on $\Delta$. In terms of both computation time and accuracy examined in Fig.~\ref{Calc_time}(a), it is reasonable to choose $\Delta$ between $5^{\circ}$ and $10^{\circ}$, for which the triple decomposition can be applied with a moderate computation time even for N512. \par
Figure~\ref{Calc_time}(b) shows the PDFs of $\omega/\langle\omega\rangle$, $\omega_{\mathrm{RR}}/\langle\omega\rangle$, and $\omega_{\mathrm{SH}}/\langle\omega\rangle$, while Fig.~\ref{Calc_time}(c) shows the PDFs of $s/\langle s\rangle$, $s_{\mathrm{EL}}/\langle s\rangle$, and $s_{\mathrm{SH}}/\langle s\rangle$. Here, the triple decomposition is applied with $\Delta=2^{\circ}$ or $5^{\circ}$ for N128. The PDFs do not change with $\Delta$ for $\Delta=2^{\circ}$ and $5^{\circ}$. Figures~\ref{Calc_time}(b,c) are further discussed in the next subsection. 
\par
\begin{figure}[t]
  \begin{center}
    \includegraphics[width=1\linewidth,keepaspectratio]{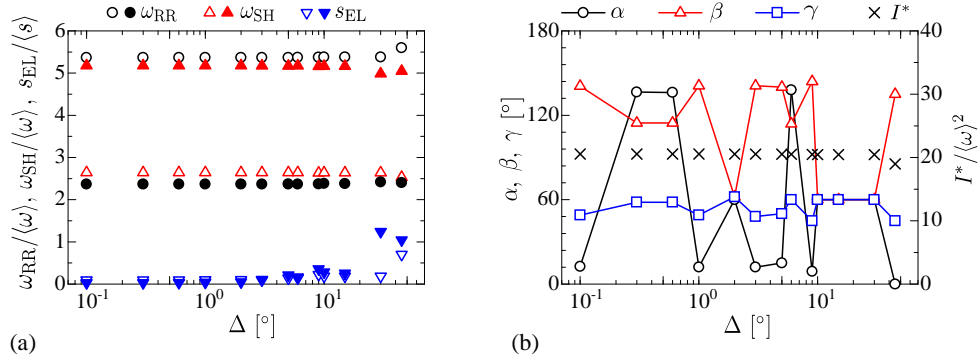}
    \caption{(a) Dependence of $\omega_{\mathrm{RR}}$, $\omega_{\mathrm{SH}}$, and $s_{\mathrm{EL}}$ on $\Delta$ at a flow point with the maximum value of $\omega$ (open symbols) and the maximum value of $s$ (closed symbols) in N128. (b) Maximum interaction scalar $I^{*}$ and angles $(\alpha,\beta,\gamma)$ that define the basic reference frame for various values of $\Delta$ at the location with the maximum value of $\omega$ in the flow (N128). }
    \label{dAngle_RR_SH_EL}
  \end{center}
\end{figure}
The effect of $\Delta$ is further examined by applying the triple decomposition to the velocity gradient tensor at a point in the flow for $0.1^{\circ}\leq\Delta\leq45^{\circ}$ in N128. Values of $\omega_{\mathrm{RR}}$, $\omega_{\mathrm{SH}}$, and $s_{\mathrm{EL}}$ are plotted against $\Delta$ in Fig.~\ref{dAngle_RR_SH_EL}(a), where the triple decomposition is applied at a flow point with the maximum value of $\omega$ or the maximum value of $s$. $s_{\mathrm{EL}}$ is negligible compared with  $\omega_{\mathrm{RR}}$ and $\omega_{\mathrm{SH}}$ at these locations. When $\Delta$ is not extremely large, $\Delta$ hardly affects $\omega_{\mathrm{RR}}$, $\omega_{\mathrm{SH}}$, and $s_{\mathrm{EL}}$. \par
Figure~\ref{dAngle_RR_SH_EL}(b) shows the angles $(\alpha,\beta,\gamma)$ used for computing the basic reference frame. The largest interaction scalar $I^{*}$ for $(\alpha,\beta,\gamma)$ is also shown in the figure. These results are plotted as a function of $\Delta$. $(\alpha,\beta,\gamma)$ of the detected basic reference frame changes with $\Delta$. However, the interaction scalar computed for $(\alpha,\beta,\gamma)$ does not change with $\Delta$. Thus, more than one set of $(\alpha,\beta,\gamma)$ exists that yield the maximum value of the interaction scalar. The interaction scalar is closely related to the results of the triple decomposition because it is related to the extracted shear tensor.~\cite{kolavr2007vortex} The independence of the maximum value of the interaction scalar from $\Delta$ explains why the results of the triple decomposition $\omega_{\mathrm{RR}}$, $\omega_{\mathrm{SH}}$, and $s_{\mathrm{EL}}$ depend little on $\Delta$. It is noteworthy that a similar observation was performed for general three-dimensional tensor data,~\cite{kolavr2007vortex} where the same results of the triple decomposition can be obtained for different basic reference frames. This is further supported by the present results for the velocity gradient tensor in homogeneous isotropic turbulence. 
\par
\subsection{Irrotational straining motion, rigid-body rotation, and shearing motion extracted with triple decomposition}
The PDFs shown in Figs.~\ref{Calc_time}(b,c) are further discussed here. We have also confirmed that the shape of the PDFs is similar for N128 and N512 although the results in N512 are not shown in the figures. In Figs.~\ref{Calc_time}(b,c), $\langle \omega\rangle$ and $\langle s\rangle$ are used for normalization for comparing magnitude between $\omega_{\mathrm{RR}}$ and $\omega_{\mathrm{SH}}$ and between $s_{\mathrm{EL}}$ and $s_{\mathrm{SH}}$. Because of the decomposition, $\omega_{\mathrm{RR}}$ and $\omega_{\mathrm{SH}}$ are smaller than $\omega$. A large probability for moderate values of $\omega/\langle\omega\rangle\approx 0.6$ can be found, while the PDFs for larger values of $\omega/\langle\omega\rangle$ exhibit nonzero values because of an intermittent profile of the high vorticity region. After the decomposition, the PDFs exhibit vastly different profiles for $\omega_{\mathrm{RR}}$ and $\omega_{\mathrm{SH}}$. The PDF of $\omega_{\mathrm{RR}}$ increases as $\omega_{\mathrm{RR}}$ becomes smaller, whereas the PDF of $\omega_{\mathrm{SH}}$ exhibits a peak for $\omega_{\mathrm{SH}}/\langle\omega\rangle\approx 0.6$, similarly to $\omega$. These PDFs indicate that flow regions with a moderate level of vorticity $\omega/\langle\omega\rangle\approx0.6$ are primarily associated with a shearing motion extracted as $\omega_{\mathrm{SH}}$. Similarly to $\omega$, $\omega_{\mathrm{RR}}$, and $\omega_{\mathrm{SH}}$, the PDF of $s_{\mathrm{EL}}$ increases with decreasing $s_{\mathrm{EL}}$, while the PDFs of $s$ and $s_{\mathrm{SH}}$ exhibit large values at their moderate values ($s/\langle s \rangle\approx 0.8$ and $s_{\mathrm{SH}}/\langle s \rangle\approx 0.5$). The flow primarily exhibits a weak irrotational straining motion ($s_{\mathrm{EL}}\approx0$) and a moderately strong shearing motion ($s_{\mathrm{SH}}>0$). \par
\begin{figure}[t]
  \begin{center}
    \includegraphics[width=0.9\linewidth,keepaspectratio]{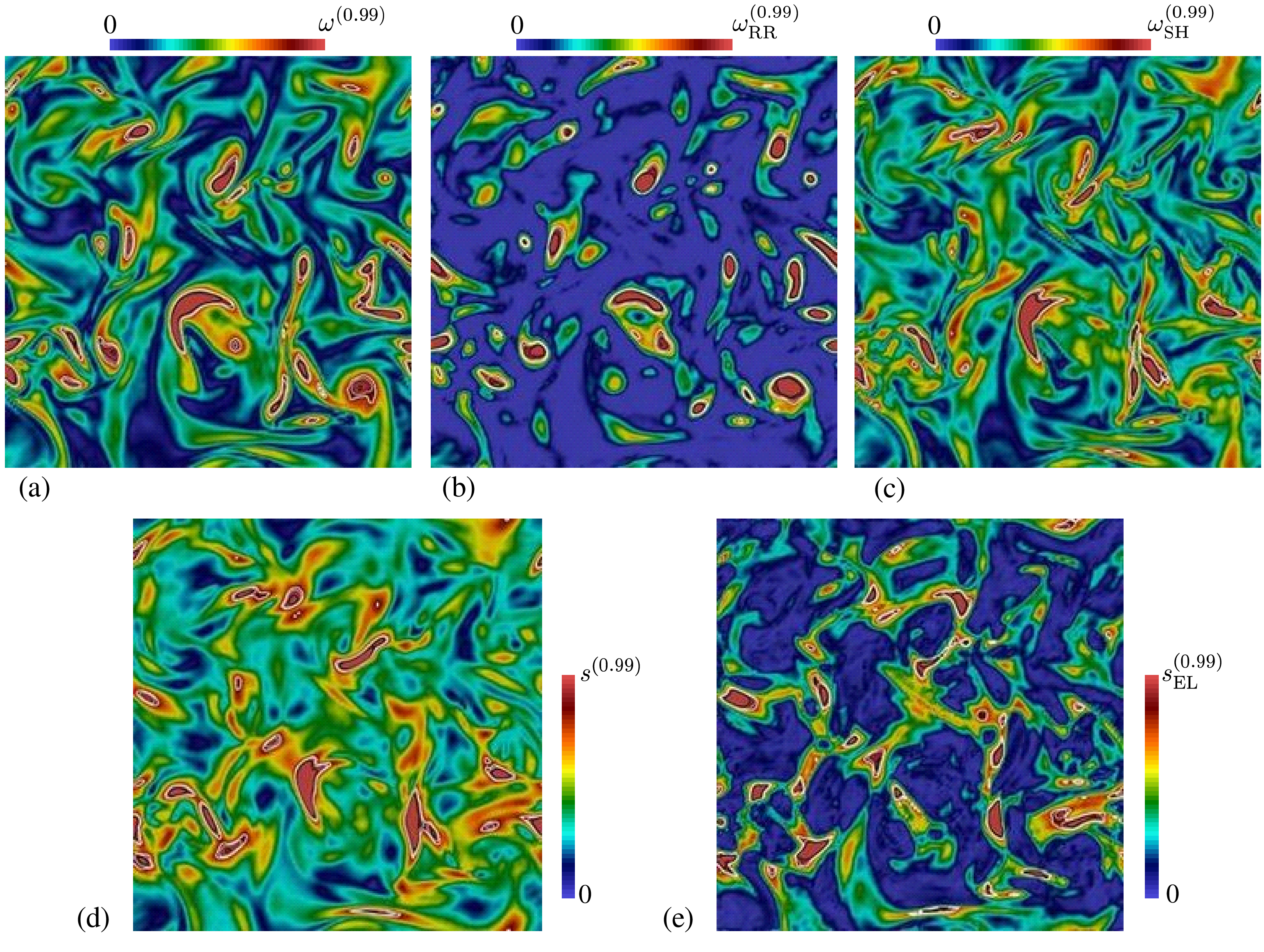}
    \caption{Color contour of (a) $\omega$, (b) $\omega_{\mathrm{RR}}$, (c) $\omega_{\mathrm{SH}}$, (d) $s$, and (e) $s_{\mathrm{EL}}$  on a $x$-$y$ plane in N128, where the triple decomposition is applied with $\Delta = 2^{\circ}$. White, gray, and black lines represent isolines of $f=f^{(0.97)}$,  $f=f^{(0.98)}$, and $f=f^{(0.99)}$, respectively. }
    \label{Fig_2D_N128}
  \end{center}
\end{figure}
For a variable $\phi$ with the PDF $p(\phi)$, a cumulative distribution function $P(\psi)$ can be defined as $P(\psi)=\int_{-\infty}^{\psi}p(\phi)d\phi$, where $P(\psi)$ represents the probability of $\phi<\psi$. Herein, a value of $\psi$ that satisfies $P(\psi)=c$ is denoted with superscript $(c)$. Figure~\ref{Fig_2D_N128} visualizes $\omega$, $\omega_{\mathrm{RR}}$, $\omega_{\mathrm{SH}}$, $s$, and $s_{\mathrm{EL}}$  with isolines of $f=f^{(0.97)}$, $f=f^{(0.98)}$, and $f=f^{(0.99)}$ on a two-dimensional plane in N128. The flow primarily exhibits $\omega_{\mathrm{RR}}\approx0$, and a region with a large $\omega_{\mathrm{RR}}$ is highly intermittent. A region with large $\omega_{\mathrm{RR}}$  can be considered as a vortex with a strong rigid-body rotation. Some isolated regions with large $\omega_{\mathrm{RR}}$ exhibit a circular shape, as shown in Fig.~\ref{Fig_2D_N128}(b), where a slice of the vortex is visualized on a two-dimensional plane. These vortices are connected with each other by regions with moderately large $\omega$ when they are visualized with $\omega$ in Fig.~\ref{Fig_2D_N128}(a). In Fig.~\ref{Fig_2D_N128}(c), shearing motions with moderately large $\omega_{\mathrm{SH}}$ exist primarily in the flow. Comparing Figs.~\ref{Fig_2D_N128}(a) and (c), some patterns in the color contour of $\omega_{\mathrm{SH}}$ are similar to those of $\omega$. This indicates that these regions visualized with $\omega$ are associated with the shearing motion rather than the rigid-body rotation. In Figs.~\ref{Fig_2D_N128}(c) and \ref{Fig_2D_N128}(d), similar patterns can be found in the regions with large $\omega_{\mathrm{SH}}$ and $s$, where the shearing motions identified by the triple decomposition are related to the rate-of-strain tensor. However, regions with large $s$ or large $\omega$ are not always coincident with regions with large $\omega_{\mathrm{SH}}$ because large $s$ and $\omega$ can also be related to large $s_{\mathrm{EL}}$ and $\omega_{\mathrm{RR}}$, respectively. Thus, Fig.~\ref{Fig_2D_N128} confirms that three different motions considered in the triple decomposition cannot be distinguished simply by $s$ and $\omega$ in the double decomposition. \par
\begin{table}[t]
  \begin{center}
  \caption{Values of $f^{(0.98)}/\langle f\rangle$ and $f^{(0.98)}/f^{(2.9\mathrm{rms})}$ for $f=\omega_{\mathrm{SH}}$, $\omega_{\mathrm{RR}}$, and $s_{\mathrm{EL}}$. }
  \label{Table_V098}
  \begingroup
\scalebox{0.85}{ 
  \begin{tabular}{ccccccccccccccc}
  \hline
  Case
  & $\omega_{\mathrm{SH}}^{(0.98)}/\langle \omega_{\mathrm{SH}} \rangle$
  & $\omega_{\mathrm{RR}}^{(0.98)}/\langle \omega_{\mathrm{RR}} \rangle$
  & $s_{\mathrm{EL}}^{(0.98)}     /\langle s_{\mathrm{EL}}      \rangle$
  & $\omega_{\mathrm{SH}}^{(0.98)}/\omega_{\mathrm{SH}}^{(\mathrm{2.9rms})}$
  & $\omega_{\mathrm{RR}}^{(0.98)}/\omega_{\mathrm{RR}}^{(\mathrm{2.9rms})}$
  & $s_{\mathrm{EL}}^{(0.98)}     /s_{\mathrm{EL}}^{(\mathrm{2.9rms})}     $
  \\
  \hline

  N128
  & $2.41$
  & $6.00$
  & $3.75$
  & $1.05$
  & $0.95$
  & $0.97$\\
  N512
  & $2.88$
  & $7.00$
  & $3.99$
  & $1.06$
  & $0.98$
  & $0.99$\\
  \hline
  \end{tabular} 
  }
  \endgroup
  \end{center}
\end{table}
\begin{figure}[t]
  \begin{center}
    \includegraphics[width=1.0\linewidth,keepaspectratio]{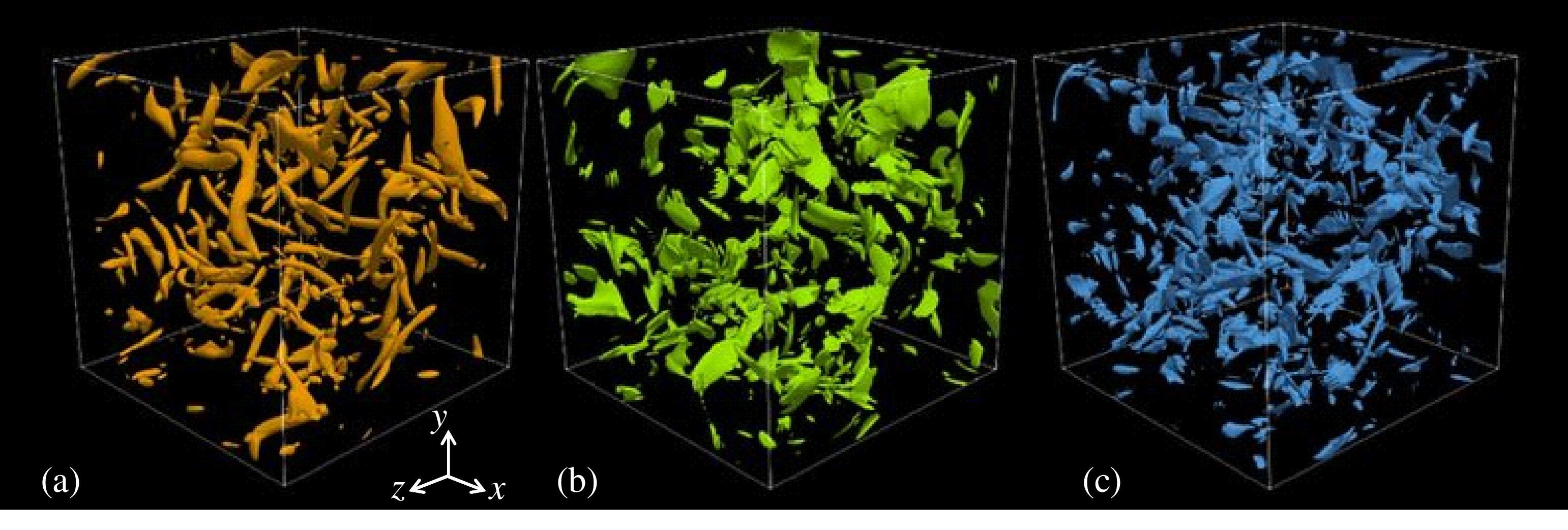}
    \caption{Three-dimensional visualization of isosurfaces of (a) $\omega_{\mathrm{RR}}=\omega_{\mathrm{RR}}^{(0.98)}$, (b) $\omega_{\mathrm{SH}}=\omega_{\mathrm{SH}}^{(0.98)}$, and (c)  $s_{\mathrm{EL}}=s_{\mathrm{EL}}^{(0.98)}$ in N128. }
    \label{3D_TDM}
  \end{center}
\end{figure}
Hereafter, as examples of flow structures detected by the triple decomposition, $f^{(0.98)}$ is used as the level of the isosurfaces of variable $f$ in three-dimensional visualization. Hence, the region surrounded by the isosurfaces occupies 2\% of the computational domain. This method that is based on the cumulative distribution function has been used for choosing the threshold for detecting small-scale vortices in turbulence.~\cite{da2011intense,jahanbakhshi2015baroclinic,jimenez1993structure,kida1998identification} Other previous studies visualize isosurfaces whose values are determined based on mean values of variables.~\cite{yeung2015extreme,buaria2019extreme} Table~\ref{Table_V098} shows $\omega_{\mathrm{RR}}^{(0.98)}$,  $\omega_{\mathrm{SH}}^{(0.98)}$,  and $s_{\mathrm{EL}}^{(0.98)}$ divided by the average of the corresponding variables. The values corresponding to the cumulative probability of 0.98 divided by the average become larger as $Re$ increases. $\omega_{\mathrm{RR}}^{(0.98)}/\langle \omega_{\mathrm{RR}}\rangle\approx6$-7 is larger than the other quantities because of the large probability of $\omega_{\mathrm{RR}}=0$. Isosurface values in flow visualization can also be determined based on a mean value and rms fluctuation~\cite{kaneda2006high} as $f^{(\alpha\mathrm{rms})}=\langle f\rangle+\alpha f_{\mathrm{rms}}$. We have calculated $f^{(\alpha\mathrm{rms})}$ for a wide range of $\alpha$, and found that $f^{(0.98)}$ is very close to $f^{(2.9\mathrm{rms})}$. Table~\ref{Table_V098} shows $f^{(0.98)}/f^{(2.9\mathrm{rms})}$, which is close to 1 for all cases. Therefore, isosurfaces of $f=f^{(0.98)}$ mostly visualize flow regions with $f\geq \langle f\rangle+2.9f_{\mathrm{rms}}$. It is also noteworthy that values of $f^{(0.98)}/f^{(2.9\mathrm{rms})}$ hardly depend on the Reynolds number considered in this study. From the definition, the region surrounded by the isosurface becomes smaller as the value of the isosurface is increased. However, most isolines of $f=f^{(0.97)}$, $f=f^{(0.98)}$, and $f=f^{(0.99)}$ surround the same regions in Fig.~\ref{Fig_2D_N128}, and the isosurface of $f=f^{(0.97)}$ and $f=f^{(0.99)}$ in three-dimensional visualization leads to the same observations described below. \par
Figure~\ref{3D_TDM} shows the isosurfaces of $\omega_{\mathrm{RR}}=\omega_{\mathrm{RR}}^{(0.98)}$, $\omega_{\mathrm{SH}}=\omega_{\mathrm{SH}}^{(0.98)}$,  and $s_{\mathrm{EL}}=s_{\mathrm{EL}}^{(0.98)}$. Tube-like vortices and sheet-like structures with large $\omega$ have been reported in the visualization of $\omega$ or $\omega^2$ in previous studies.~\cite{dubief2000coherent} The sheet-like structures do not appear in the visualization of $\omega_{\mathrm{RR}}$ in Fig.~\ref{3D_TDM}(a), where tube-like vortices can be clearly observed. In contrast, a large number of sheet-like structures are shown in the region with strong shear in the visualization of $\omega_{\mathrm{SH}}$, as shown in Fig.~\ref{3D_TDM}(b). In the triple decomposition, strong rigid-body rotation and shearing motion can be detected by thresholding $\omega_{\mathrm{RR}}$ and $\omega_{\mathrm{SH}}$, respectively. The isosurface of $s_{\mathrm{EL}}=s_{\mathrm{EL}}^{(0.98)}$ in Fig.~\ref{3D_TDM}(c) is very intermittent, and each isolated region with $s_{\mathrm{EL}}=s_{\mathrm{EL}}^{(0.98)}$ occupies a smaller volume than the isosurface of other quantities. It is noteworthy that $s_{\mathrm{EL}}$ exhibits a large probability for $s_{\mathrm{EL}}\approx 0$ in Fig.~\ref{Calc_time}(c), and the irrotational straining motion $s_{\mathrm{EL}}$ is weaker than the shearing motion $\omega_{\mathrm{SH}}$. \par
\begin{figure}[t]
  \begin{center}
    \includegraphics[width=1\linewidth,keepaspectratio]{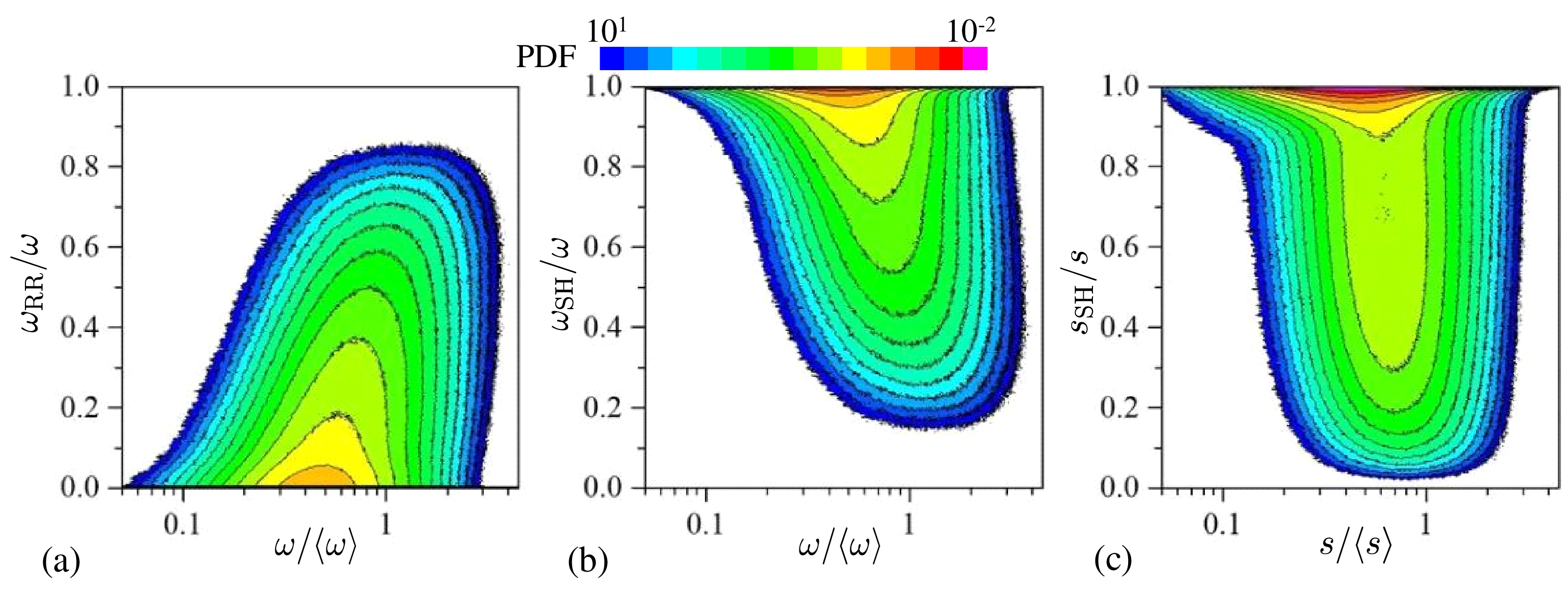}
    \caption{Joint PDFs of (a) $\omega$ and $\omega_{\mathrm{RR}}/\omega$, (b) $\omega$ and $\omega_{\mathrm{SH}}/\omega$, and (c) $s$ and $s_{\mathrm{SH}}/s$. The results are obtained with triple decomposition with $\Delta = 5^{\circ}$ from N512. }
    \label{JPDF_OS_TDM}
  \end{center}
\end{figure}
Figures~\ref{JPDF_OS_TDM}(a) and (b) show the joint PDFs between $\omega$ and $\omega_{\mathrm{RR}}/\omega$ and between $\omega$ and $\omega_{\mathrm{SH}}/\omega$, respectively. From the definition, $\omega_{\mathrm{RR}}$ and $\omega_{\mathrm{SH}}$ satisfy $\omega\geq\omega_{\mathrm{RR}}$ and $\omega\geq\omega_{\mathrm{SH}}$, respectively. A large probability appears for $\omega_{\mathrm{RR}}/\omega\approx0$ and $\omega_{\mathrm{SH}}/\omega\approx1$, and the flow primarily exhibits shearing motions without a strong rigid-body rotation. Positive and negative correlations with $\omega$ exist for $\omega_{\mathrm{RR}}/\omega$ and $\omega_{\mathrm{SH}}/\omega$, respectively. A larger $\omega_{\mathrm{RR}}/\omega$ and a smaller $\omega_{\mathrm{SH}}/\omega$ are more frequently found for a larger $\omega$. Thus, the relative importance of the rigid-body rotation and shearing motion, evaluated as $\omega_{\mathrm{RR}}/\omega$ and $\omega_{\mathrm{SH}}/\omega$, respectively, depends on $\omega$. The rigid-body rotation becomes more important for the region with large $\omega$, while the shearing motion dominates over the rigid-body rotation for the region with small $\omega$. This tendency agrees with the visualization of $\omega_{\mathrm{SH}}$ and $\omega$ in Fig.~\ref{Fig_2D_N128}, where $\omega_{\mathrm{SH}}$ exhibits nonzero values even for small values of $\omega$. Figure~\ref{JPDF_OS_TDM}(c) shows the joint PDFs between $s$ and $s_{\mathrm{SH}}/s$. Flow regions with large $s$ do not always correspond to large $s_{\mathrm{SH}}/s$ even though large PDF values appear for $s_{\mathrm{SH}}/s\approx1$ for a wide range of $s$. The joint PDF and visualization in Fig.~\ref{Fig_2D_N128} confirm that strength of the shearing motion cannot be accurately quantified by $s$. \par
\begin{figure}[t]
  \begin{center}
    \includegraphics[width=1\linewidth,keepaspectratio]{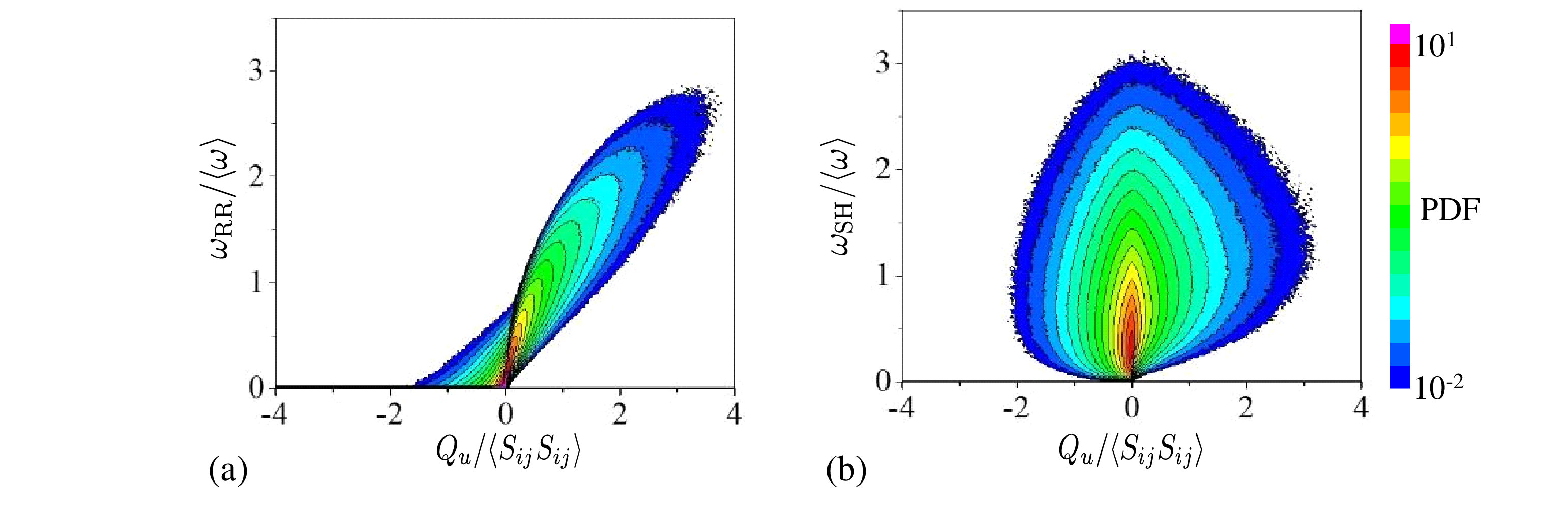}
    \caption{(a) Joint PDF of $\omega_{\mathrm{RR}}$ and $Q_{u}$. (b) Joint PDF of $\omega_{\mathrm{SH}}$ and $Q_{u}$. The results are obtained with triple decomposition with $\Delta = 5^{\circ}$ from N512. }
    \label{JPDF_Qu_OmRR_rev}
  \end{center}
\end{figure}
Figure~\ref{JPDF_Qu_OmRR_rev}(a) shows the joint PDF of $\omega_{\mathrm{RR}}/\langle \omega\rangle$ and $Q_{u}/\langle S_{ij}S_{ij}\rangle$ in N128. In the double decomposition, small-scale vortices are often detected as the region with $Q_u>0$. The joint PDF shows that the region with $Q_u>0$ tends to exhibit $\omega_{\mathrm{RR}}/\langle \omega\rangle>0$. Meanwhile, a large probability of $\omega_{\mathrm{RR}}/\langle \omega\rangle\approx0$ is found for $Q_u<0$. The joint PDF indicates that the strong rigid-body rotation with large $\omega_{\mathrm{RR}}$ can be detected by $Q_u$. However, a flow region with $Q_u\approx0$ can exhibit a moderately large value of $\omega_{\mathrm{RR}}$, which is difficult to detect with $Q_u$. Figure~\ref{JPDF_Qu_OmRR_rev}(b) shows the joint PDF of $\omega_{\mathrm{SH}}/\langle \omega\rangle$ and $Q_{u}/\langle S_{ij}S_{ij}\rangle$. Large $\omega_{\mathrm{SH}}$ tends to appear for $Q_{u}\approx0$, where $s$ is as large as $\omega$. This result agrees well with previous studies of vortex sheets, for which the enstrophy is comparable to the strain rate.~\cite{tanaka1993characterization} It should be noted that the triple decomposition can detect regions with strong shearing motions solely with $\omega_{\mathrm{SH}}$ while most methods proposed for detecting vortex sheets use more than one variable.~\cite{tanaka1993characterization} \par
\begin{figure}[t]
  \begin{center}
    \includegraphics[width=0.90\linewidth,keepaspectratio]{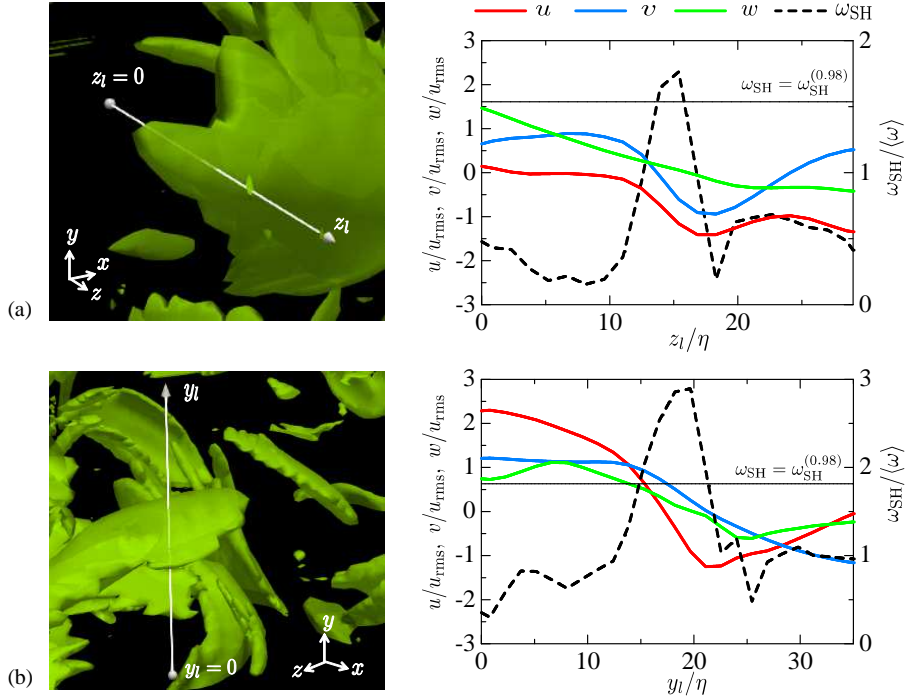}
    \caption{Profile of velocity components $(u,v,w)$ and $\omega_{\mathrm{SH}}$ on the local coordinate (shown with an arrow in left figures) across the strong shear region visualized with the isosurface of $\omega_{\rm SH} = \omega^{\rm (0.98)}_{\rm SH}$ (green). (a) N128 and (b) N512. The local coordinates $y_{l}$ and $z_{l}$ align with the $y$ and $z$ directions, respectively. Horizontal line represents $\omega_{\rm SH} = \omega^{\rm (0.98)}_{\rm SH}$ in the right figures. }
    \label{Plot_over_line}
  \end{center}
\end{figure}
An important application of the triple decomposition is the detection of the internal shear layer in turbulent flows. In previous studies, the layer-like structures in turbulence have also been observed in visualization of $s$ or $\omega$. However, regions detected by thresholding $\omega$ can be associated with rigid-body rotation rather than the shearing motion in Fig.~\ref{Fig_2D_N128}. Similarly, regions with large $s$ are not always related to shear layers because of $s_{\mathrm{EL}}$. Therefore, the detection of shearing motions with the double decomposition requires combining $s$ or $\omega$ with other criteria. For example, the shear layer might be detected by visually assessing the shape of isosurface of $\omega$. The advantage of the triple decomposition in the shear layer detection is that strength of the shearing motion is quantified solely by $\omega_{\mathrm{SH}}$ defined at each point in a flow. Then, it is expected that the shear layers can be detected simply as regions with large $\omega_{\mathrm{SH}}$. \par
Figure~\ref{Plot_over_line} shows sheet-like structures of the strong shear region with $\omega_{\rm SH} \geq \omega^{\rm (0.98)}_{\rm SH}$ in (a) N128 and (b) N512, where the normal direction of the shear layers is close to the $z$ and $y$ directions, respectively. Additionally, the figures show the profiles of $u$, $v$, $w$, and $\omega_{\mathrm{SH}}$ on the local coordinate $y_{l}$ or $z_{l}$ normalized by the Kolmogorov length scale $\eta$, where $y_{l}$ and $z_{l}$ align with the $y$ and $z$ directions, respectively. Components of $(\nabla\bd{u})_{\rm SH}$ can be used for examining the direction of detected shear layers although this is not addressed in this study. On the local coordinate, $\omega_{\mathrm{SH}}$ increases rapidly, reaches a peak exceeding $\omega_{\mathrm{SH}}^{(0.98)}$, and subsequently decreases rapidly. In Figs.~\ref{Plot_over_line}(a) and (b), the thickness of the large $\omega_{\mathrm{SH}}$ region can be estimated as $10\eta$ for both Reynolds numbers. In Fig.~\ref{Plot_over_line}(a), two velocity components $u$ and $v$ that are parallel to the strong shear region change significantly across the shear layer. The difference in $u$ and $v$ between two sides of the shear layer is close to the rms velocity fluctuation $u_{\mathrm{rms}}$. A similar velocity jump across the shear layer can be found for $u$ for the high Reynolds number case in Fig.~\ref{Plot_over_line}(b), where $u$ changes by more than $u_{\mathrm{rms}}$. The velocity jump  and the thickness of the detected shear layer can be related to the extreme events in turbulence whose characteristic length and velocity scales are $\eta$ and  $u_{rms}$, respectively.~\cite{jimenez1993structure,buaria2019extreme} An average velocity field extracted in the strain eigenframe also exhibited a shear layer whose thickness scales with the Kolmogorov scale,~\cite{elsinga2017scaling} while the present analysis identifies the shear layer in an instantaneous three-dimensional velocity field. The thickness and velocity jump across the strong shear layer are similar for different Reynolds numbers in N128 and N512 when they are normalized by $u_{\mathrm{rms}}$ and $\eta$. The rms velocity fluctuation is often considered as the characteristic velocity scale of large-scale motions in turbulent flows. The velocity jump of the order of $u_{\mathrm{rms}}$ across the shear layer implies that the strong shear layer detected by the triple decomposition separates regions of different large-scale motions. Sheet-like structures are found in the visualization of isosurface of vorticity magnitude in homogeneous isotropic turbulence.~\cite{jimenez1993structure} These structures are often called vortex sheets. It was found that the thickness of the vortex sheets was related to the Kolmogorov scale, while the velocity difference across the sheet was close to the rms velocity fluctuation. These characteristics of the vortex sheets agree with the shear layer detected with $\omega_{\mathrm{SH}}$. This implies that the vortex sheets found with the double decomposition are related to the shearing motion with large $\omega_{\mathrm{SH}}$. \par
\section{Concluding Remarks}\label{Sec_Concluding}
A detailed procedure was proposed for the triple decomposition~\cite{kolavr2007vortex} in three-dimensional flows. This procedure comprised two parts: the identification of the basic reference frame from all reference frames obtained by three sequential rotational transformations defined with angles $(\alpha, \beta, \gamma)$ applied to the Cartesian coordinate; the decomposition of the velocity gradient tensor in the basic reference frame into three components associated with an irrotational straining motion, rigid-body rotation, and shearing motion. Here, the basic reference frame, defined as the reference frame with the largest interaction scalar, was chosen from a finite number of reference frames by discretely setting $(\alpha, \beta, \gamma)$ with the angle interval $\Delta$. \par
The triple decomposition of the velocity gradient tensor was tested for the DNS database of an incompressible, forced homogeneous isotropic turbulence with the turbulent Reynolds numbers $Re_{\lambda}=27$ and 140. It was shown that more than one basic reference frame could provide the largest interaction scalar, whose value depended little on the specific choice of the basic reference frame. The results of the triple decomposition were shown to be robust for a wide range of $\Delta$. The computation time of the triple decomposition changed as $(180^{\circ}/\Delta+1)^2 (90^{\circ}/\Delta+1)$. In terms of the $\Delta$ dependences of the triple decomposition and its computation time, $\Delta$ between $2^{\circ}$ and $5^{\circ}$ was reasonable for applying the triple decomposition to the three-dimensional data of the velocity gradient tensor in turbulent flows. \par
The results of the triple decomposition in homogeneous isotropic turbulence were presented with $\omega_{\mathrm{SH}}$, $\omega_{\mathrm{RR}}$, and $s_{\mathrm{EL}}$, representing the shearing motion, rigid-body rotation, and irrotational straining motions, respectively. Regions with strong rigid-body rotations or irrotational strains (large $\omega_{\mathrm{RR}}$ or $s_{\mathrm{EL}}$) appeared intermittently in space. Most flow regions exhibited small $\omega_{\mathrm{RR}}$ and $s_{\mathrm{EL}}$ while shearing motions with moderately large $\omega_{\mathrm{SH}}$ appeared in most flow regions. The flow visualized with large $\omega_{\mathrm{RR}}$ exhibited a tube-like structure of vortices. A strong shear region (large $\omega_{\mathrm{SH}}$) exhibited sheet-like structures (shear layers). Because a region with large $\omega$ or large $s$ did not always possess large $\omega_{\mathrm{SH}}$, the strong shear region could not be simply detected by thresholding $\omega$ or $s$. The region with large $\omega_{\mathrm{SH}}$ often appeared when $\omega$ was comparable to $s$, i.e. $Q_u\approx0$. The velocity components parallel to the shear layer exhibited a sharp jump within the shear layer, where the velocity difference across the shear layer is of the order of the rms velocity fluctuation and the layer thickness was approximately $10\eta$. \par
\section*{Acknowledgment}
This work was partially supported by the Collaborative Research Project on Computer Science with High-Performance Computing in Nagoya University and by JSPS KAKENHI Grant Number 18K13682 and 18H01367. We would like to thank Editage (www.editage.jp) for English language editing. 
%




\bibliographystyle{elsarticle-num}
\bibliography{bibtex_watanabe}







\end{document}